\documentclass[twocolumn]{aastex63} 

\usepackage[colorinlistoftodos]{todonotes}
\usepackage[flushleft]{threeparttable}
\usepackage{graphicx}
\usepackage{tabularx}

\let\Oldtodo\todo
\renewcommand{\todo}[1]{\Oldtodo[inline]{#1}}

\hypersetup{linkcolor=cyan,citecolor=cyan,filecolor=cyan,urlcolor=cyan}

\usepackage{amsmath}
\usepackage{float}
\usepackage{cleveref}

\shorttitle{Multis Through hot Jupiter Destruction}
\shortauthors{Liveoak, Hallatt \& Millholland}

\graphicspath{{./}{figures/}}

\begin{document}

\title{Stability of Multiplanet Systems Through Hot Jupiter Destruction}

\correspondingauthor{Donald Liveoak}
\email{dliveoak@umich.edu}

\author[0000-0003-4992-8427]{Donald Liveoak}
\affil{Department of Physics, University of Michigan, Ann Arbor, MI 48109}
\affil{MIT Kavli Institute for Astrophysics and Space Research, Massachusetts Institute of Technology, Cambridge, MA 02139, USA}

\author[0000-0003-4992-8427]{Tim Hallatt}
\affil{MIT Kavli Institute for Astrophysics and Space Research, Massachusetts Institute of Technology, Cambridge, MA 02139, USA}

\author[0000-0003-4992-8427]{Sarah Millholland}
\affil{MIT Kavli Institute for Astrophysics and Space Research, Massachusetts Institute of Technology, Cambridge, MA 02139, USA}

\begin{abstract}
Recent observational and theoretical work suggests that the sub-Jovian desert (periods ${\lesssim}3$ days, masses ${\sim}10{-}100 \ M_{\oplus}$) hosts the remains of destroyed hot Jupiters (``desert dwellers"). In this work, we explore how differing hot Jupiter destruction mechanisms -- Roche lobe overflow (RLO) vs. tidal disruption during high eccentricity migration (HEM) -- may be discerned observationally based on the presence of companion planets to desert dwellers. We show that gas giant destruction via RLO clears out the desert of any companions inside orbital periods ${\lesssim}$4 days; desert dwellers should sit alone in the desert if they form through this mechanism. Numerically mapping the instability threshold in planet mass and orbital distance, we find that the majority of observed companions to desert dwellers are safely in the stability region. RLO therefore does not preclude the existence of nearby companions beyond the desert, in contrast to gas giant tidal disruption during HEM. Further characterization of desert dweller systems may therefore elucidate the fates of hot Jupiters.

\end{abstract}

\section{Introduction} \label{sec:intro}

The distribution of the shortest-period exoplanets is roughly bimodal: hot Jupiters (``HJs"; ${\gtrsim}10 \ R_{\oplus}$, ${\gtrsim}100 \ M_{\oplus}$) and super-Earths/sub-Neptunes (${\lesssim}4 \ R_{\oplus}$, ${\lesssim} 10 \ M_{\oplus}$) inside orbital periods ${\lesssim}$3 days are separated by a gulf nearly devoid of intermediate-sized planets. This dearth of hot, mid-sized planets is the ``sub-Jovian desert" \citep[e.g.][]{szakis11, mazholfai16}. The desert and its boundaries in orbital period, planet mass, and radius provide a window into the migratory and mass loss processes sculpting the hottest exoplanets. 

Stretching upward in mass/radius toward smaller periods, the upper boundary of the desert is thought to reflect the tidal disruption and circularization distances of hot Jupiters emplaced via high eccentricity migration (HEM) \citep[][see \citealt{baibat18} however for an in-situ formation scenario]{owelai18}. Dropping in mass/radius toward shorter periods, the lower boundary is likely a remnant of atmospheric erosion of lower mass planets, initiated by intense stellar radiation at such tight distances (\cite{owelai18,hallee22}; see \cite{visknugre22} for observational evidence that the desert's upper boundary is stable to mass loss, and \cite{kurnak14} for a differing theoretical perspective). Alternatively, the lower boundary may reflect the differing Roche limit (set by the mass/radius relationship) of less massive planets undergoing HEM like their gas giant counterparts \citep{matkon16}. Until recently, the violent nature of such processes was thought to explain why the desert was virtually empty.

Contrary to previous results based on $\textit{Kepler}$ \citep[e.g.][]{beanes13,mazholfai16,lunkjealb16}, the $\textit{TESS}$ mission \citep{ricwinvan15} has recently discovered a number of planets residing squarely in the regime previously thought uninhabited; the desert is no longer empty \citep{visbeh25,cuiarmhad26}. Such ``desert dwellers" offer fresh constraints on 
the origin/evolution of close-in exoplanets surrounding and residing within the desert.

Based on desert dwellers' and hot Jupiters' shared preference for metal-rich stars, \cite{visbeh25} suggested a ``top-down" process by which desert dwellers emerge as the exposed interiors of destroyed gas giants \citep[see e.g. \citealt{barchabar05} for early exploration of this hypothesis, though it is now understood that evaporation cannot destroy hot Jupiters; see also][for additional argument in favor of the ``top-down" interpretation]{armlopadi20}. In this picture, hot Jupiters are destroyed either during their HEM or through subsequent orbital inspiral. \cite{halmil26} showed that the entire desert can indeed be backfilled via the tidal inspiral channel; orbital decay strips gas giant atmospheres via Roche lobe overflow (RLO), before orbital expansion during mass transfer deposits the denuded remains across the desert. If correct, the theory of \cite{halmil26} implies that desert dwellers provide an opportunity to directly study the exposed interiors of giant planets. It is thus imperative to observationally confirm the theory by differentiating it from other mechanisms. The purpose of this Letter is to help differentiate the \cite{halmil26} theory from others.

The chief alternative mechanism that may conceivably populate the desert is tidal disruption during HEM. Rather than tidally-driven mass transfer, planet tidal destruction during HEM is initiated via dynamical tides excited in the planet at periapse (e.g. \cite{guiramlin11,liuguilin13}; note that we will hereafter refer to this destruction mechanism as ``HEM" for brevity), leaving behind a dense remnant core. One way of differentiating the \cite{halmil26} RLO theory from HEM is through multiplanet systems. HEM is well-known to disintegrate inner planetary systems, precluding nearby companions to hot Jupiters after migration \citep[e.g.][]{musdavjoh15,bec25}. If desert dwellers form via HEM, we therefore do not expect them to have companions. On the other hand, a sizable minority of hot Jupiters $\textit{do}$ harbor nearby planets \citep[at least ${\sim}10\%$;][]{wuricwan23,shavanhua26}.\footnote{\cite{doyarmacu25} find that ${\sim}26\%$ of planets in the desert reside in multiplanetary systems, which exceeds the estimate for hot Jupiters. The fraction of desert dwellers with companions may exceed the hot Jupiter companion rate due to the fact that desert dwellers with companions reside at the bottom boundary of the desert. Their ambiguous ``desert dweller'' status suggests a mix of planets descended from hot Jupiters as envisaged by \cite{halmil26} and planets simply comprising the upper end of the sub-Neptune population.}
If desert dwellers descend from hot Jupiters destroyed via orbital inspiral, we may therefore expect a fraction of them to reside in multiplanetary systems similarly to hot Jupiters. The question addressed in this work is thus whether multiplanetary systems remain dynamically stable through RLO of an inner hot Jupiter, or whether they are disrupted by the system's reconfiguration. As future observations continue to build the desert dweller sample, our results will help determine whether those in multiplanetary systems could have descended from hot Jupiters that underwent inspiral.\footnote{Additional conditions must of course be satisfied to argue that a specific desert dweller with outer companions likely formed via hot Jupiter RLO. These include, but are not limited to, stability of the system $\textit{before}$ tidal inspiral, high density and intermediate mass similar to the existing desert dweller sample \citep[e.g.][]{visbeh25,doyarmacu25,halmil26}, and stellar age and spectral type that allow for tidal decay. Our study's goal is more modest, aimed solely at the question of whether companions to desert dwellers that form via RLO would remain stable.}

This Letter is organized as follows. In Section \ref{sec:methods} we outline our numerical setup and implementation of hot Jupiter RLO. Section \ref{sec:results} showcases our results on system stability, surveying parameter space in the context of real multiplanet systems containing a desert dweller. We summarize avenues for future work needed to confirm desert dweller formation theories in Section \ref{sec:future}. The punchline of our investigation is provided in Section \ref{sec:conclusion}.

\section{Methods $\&$ Numerics}\label{sec:methods}

To follow the dynamics of companion planets as hot Jupiters undergo RLO, we must couple the planets' mutual gravitational interaction with hot Jupiter mass loss and orbital evolution due to mass transfer. We couple $N$-body dynamics with hot Jupiter mass transfer using \texttt{REBOUND} \citep{rein2012rebound}, augmented with \texttt{REBOUNDx} to implement processes due to mass transfer \citep{tamayo2020reboundx}. We additionally account for orbital perturbation of both planets due to stellar spin up during mass transfer (also using \texttt{REBOUNDx}). In this Section, we detail our numerical setup treating the mass transfer process in tandem with $N$-body evolution. 

\subsection{Mass Transfer}

Upon RLO, a hot Jupiter loses mass while its orbit expands. The degree of orbital expansion depends on how much angular momentum is returned to the planet's orbit. \cite{halmil26} showed that RLO can only reproduce desert dwellers if it is ``lossy": mass transfer must remove most of the hot Jupiter's orbital angular momentum as well as mass. Lossy RLO contrasts with standard mass transfer theory, in which the donor's orbital angular momentum is returned via an accretion disk that forms interior to the planet \citep[e.g.][]{ver82,priwad85}. In hot Jupiter systems, lossy RLO likely arises due to stellar accretion of angular momentum and mass, preventing the formation of a disk (Hallatt, Owen $\&$ Millholland, subm.). Without an accretion disk, the only angular momentum retained by the hot Jupiter orbit may 
arise due to material leaving the planet from the first Lagrange point possessing a lower specific angular momentum than material at the planet 
\cite[i.e. the small difference in specific angular momenta between Lagrange point and planet is retained by the orbit;][]{jiaspr17}. During lossy RLO, we therefore expect that only the mass-losing body experiences an outward torque; we only apply RLO-driven orbital expansion to the hot Jupiter, not to companions.

We implement hot Jupiter mass loss and orbital expansion using \texttt{modify\_mass} and  \texttt{modify\_orbits\_direct} in \texttt{REBOUNDx}. To prescribe an instantaneous mass loss and acceleration due to mass transfer, we employ a precomputed evolution track in planet mass $m_{\rm  p}(t)$ and semi-major axis $a(t)$ following \cite{halmil26}. This prescription amounts to a rapid decline in planet mass ($m_{\rm p}{=}250\ M_{\oplus}{\rightarrow} {\sim}m_{\rm core}$ with $m_{\rm core}$ the core mass), with $a(t)$ set by preserving a small amount of orbital angular momentum as mass is lost \cite[see Equation 19 of][]{halmil26}. To implement in $\texttt{REBOUNDx}$, we set the characteristic timescales using:

\begin{align}\label{equation:tracks}
    \tau_a &= \frac{a}{\dot{a}} = \frac{a_\text{sim}(t) \Delta t}{a_\text{track}(t+\Delta t) - a_\text{sim} (t)}\\
    \tau_{m_{\rm p}} &= \frac{m_{\rm p}}{\dot{m}_{\rm p}} = \frac{m_\text{sim}(t) \Delta t}{m_\text{track}(t+\Delta t) - m_\text{sim} (t)},
\end{align}

\noindent where $a_\text{sim}$ ($m_\text{sim}$) and $a_\text{track}$ ($m_\text{track}$) are the planet's semi-major axis (mass) in the simulation and evolution track at time $t$, respectively. Furthermore, $\Delta t$ is chosen to be small enough to ensure smooth tracking of the planet's parameters to the prescribed trajectory. Positive values of $\tau_a$ and $\tau_m$ represent exponential growth, while negative values correspond to exponential damping.

Equations \ref{equation:tracks} provide the hot Jupiter's time-evolving mass (from $\tau_{m_{\rm p}}$) and 
acceleration (through $\tau_{\rm a}$) solely due to mass transfer. We stress however that \texttt{REBOUNDx} includes $N$-body interactions simultaneously with the prescribed acceleration and mass loss; i.e. the orbit feels acceleration due to angular momentum retention during mass loss, but also due to $N$-body forces that continue to perturb the orbit.\footnote{We could equally implement RLO using \texttt{modify\_orbits\_forces} \citep[][]{tamayo2020reboundx}.} One shortcoming of our approach is that, by enforcing the hot Jupiter to undergo mass loss at the prescribed rate, we ignore possible changes in the mass loss behavior (and the coupled orbital behavior) due to orbital perturbation from the companion. Since lossy RLO is a runaway process that occurs extremely quickly (over ${\sim}10^{4}$ yr), we do not expect RLO to shut off once it initiates unless the orbit is violently perturbed, e.g. via mergers or ejections.

\subsection{Stellar Rotation}

The host stars of hot Jupiters that undergo RLO are expected to be spun up due to tidal orbital inspiral and planet-to-star mass transfer \citep[see Figure 11 of][]{halmil26}. Rapidly spinning, oblate host stars dynamically perturb planet orbits via distortion of the stellar gravity field. This oblateness-induced distortion of the stellar gravity field is quantified by the stellar quadrupole moment $J_{2}$ \citep[e.g.][]{tre23}. To account for stellar oblateness effects on both planets during spin up, we model the host star's oblateness via the \texttt{gravitational\_harmonics} effect in \texttt{REBOUNDx}, specifying the stellar quadrupole moment via \citep{spalding2016spin}:

\begin{equation}
    J_2(t) = \frac{k_2}{3}\left(\frac{\Omega(t)}{\Omega_B}\right)^2,
\end{equation}

\noindent where $k_2{=}0.028$ is the star's Love number \citep[twice the apsidal motion constant of 0.014 for a Sun-like star;][]{eggkis01}, 
$\Omega_B$ is the star's break-up spin frequency, and $\Omega(t)$ is the star's time-dependent rotation rate determined in the RLO evolution track. We approximate $\Omega_B^2{=}GM_\star /R_\star^3$, with $G$ the gravitational constant, $M_\star$ the stellar mass, and $R_\star$ the stellar radius. For simplicity, our simulations fix $M_\star{=}M_{\odot}$ and $R_\star{=}R_{\odot}$.

\begin{figure*}
    \centering
    \includegraphics[width=\linewidth]{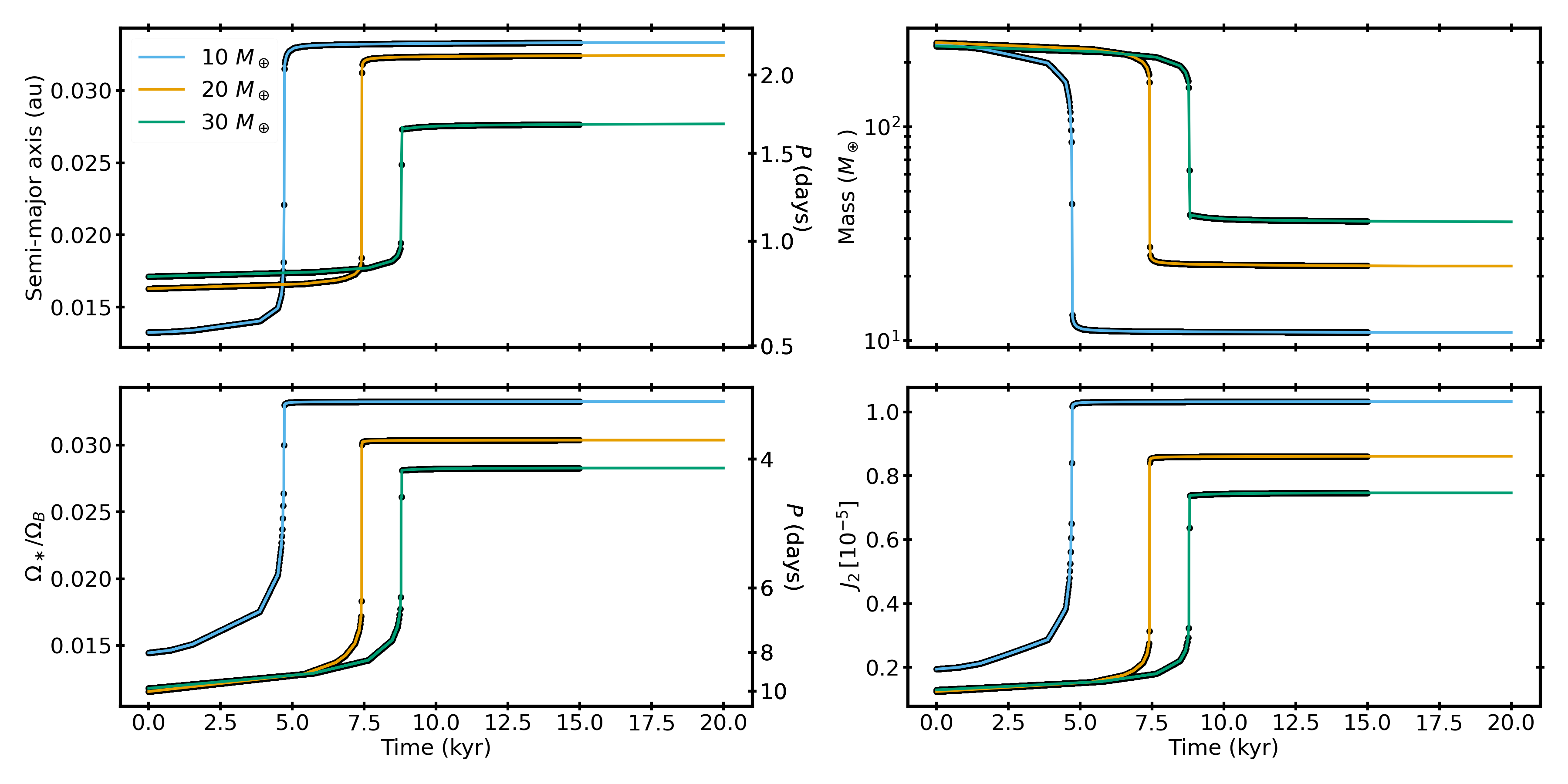}
    \caption{Hot Jupiter Roche lobe overflow evolution tracks, as computed by \cite{halmil26} (black data) and our implementation in \texttt{REBOUND} (colored data points). Blue, orange, and green data points correspond to hot Jupiters containing $10, \ 20, \mathrm{and} \ 30 \ M_{\oplus}$ cores respectively, which end mass transfer at differing periods and masses \citep[for our choice of initial planet entropy $8 \ k_{\rm B}/m_{\rm H}$ where $k_{\rm B}$ is Boltzmann's constant and $m_{\rm H}$ the hydrogen mass; see][]{halmil26}. Our hot Jupiter evolution tracks begin immediately before catastrophic mass loss to save computation time by avoiding the slow tidal decay phase ($\dot{a}{<}0$). Top left: orbital semi-major axis evolution (mapped to orbital period on right axis). Top right: mass evolution. Bottom left: stellar rotation frequency evolution (relative to breakup; mapped to rotation period on right axis). Bottom right: stellar quadrupole moment evolution. Our \texttt{REBOUND} implementation accurately reproduces the mass, stellar spin, and orbital evolution of hot Jupiters undergoing ``lossy" Roche lobe overflow.}
    \label{fig:tracking-test}
\end{figure*}

\subsection{Recapping RLO of Isolated hot Jupiters}

Before delving into the evolution of companions during RLO, we briefly recap the evolution of an isolated hot Jupiter under RLO, as shown in \Cref{fig:tracking-test}. We begin our hot Jupiter evolution tracks immediately before they undergo catastrophic mass transfer, avoiding the slow tidal decay phase ($\dot{a}{<}0$) to save computation time without affecting our results. In each case, the hot Jupiter begins with a semi-major axis $a{<}0.017$ au and a mass ${\sim}250 \ M_\oplus$. The stellar spin frequency grows as mass transfer proceeds, mainly due to tidal transfer of angular momentum from orbit to star, with lesser contribution from accretion of material onto the stellar surface. The orbit however $\textit{grows}$, as a small fraction of escaping gas' angular momentum is retained by the orbit \citep[see Equation 19 and surrounding discussion of][]{halmil26}. This orbital angular momentum return from mass loss dominates over tidal decay during this catastrophic phase of RLO. By the end of RLO, host stars can boast quadrupole moments as large as $J_{2}{\sim}10^{-5}$, comparable to that of a zero-age main sequence star \citep[see e.g. Figure 6 of][]{farnaoli23}. It is thus conceivable that stellar spin-up can perturb close-in planetary systems containing a desert dweller.

\section{Results}\label{sec:results}

\subsection{Dynamics of Single Companions During RLO}\label{subsec:singles}
We now address the evolution of companions to hot Jupiters undergoing RLO. In this section, we only consider the case of a single companion; we address multi-companion systems in Section \ref{sec:application}. We consider a broad range of companion parameters, sampling companion mass, semi-major axis, and eccentricity from $m_{\rm c}{\in}[10, 300]\ M_{\oplus}$, $a_{\rm c}{\in} [0.035,0.2]$ au, and $e{\in} [0,0.2]$ (each sampled uniformly).\footnote{We ensure that each initial condition is stable via direct $N$-body integration; see Section \ref{sec:application}.} The planets begin in a coplanar configuration but misaligned from the stellar spin axis by 5$^{\circ}$. The mean anomaly $M$, argument of periapsis $\omega$, and longitude of ascending node $\Omega$ are all sampled uniformly from $[0,2\pi)$ for both planets. We integrate for 15 kyr after RLO to probe secular dynamics and system stability. We perform three sets of simulations, each comprising 1600 integrations of the aforementioned initial conditions, with hot Jupiter core masses of $10$, $20$, and $30\,M_\oplus$.

\begin{figure*}
    \centering
    \includegraphics[width=\linewidth]{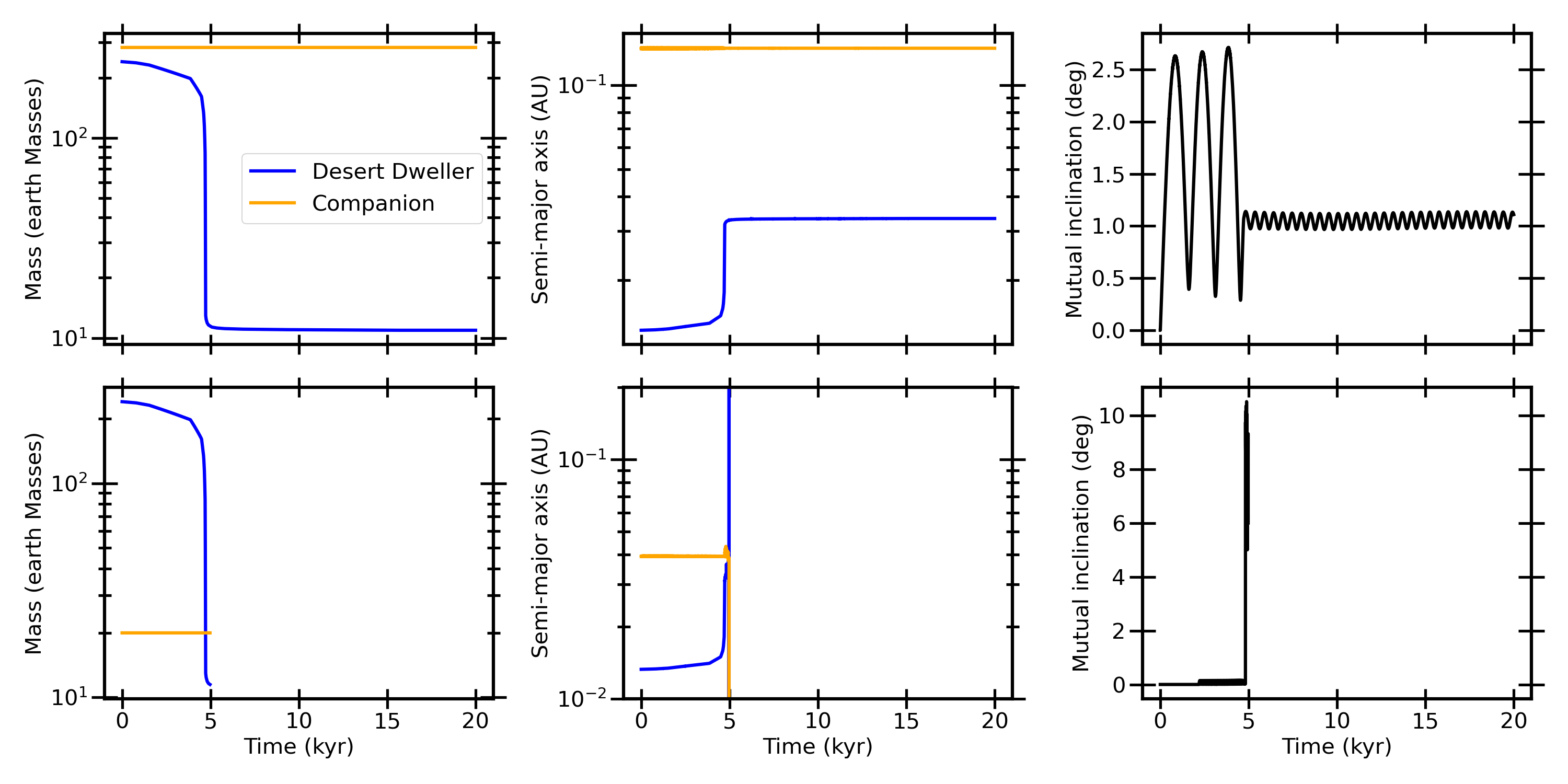}
    \caption{Two example systems (top and bottom panels) comprising an outer gas giant companion (orange curves) to an inner hot Jupiter which undergoes RLO en route to becoming a desert dweller (blue curves). Left panels depict planet mass evolution, middle planet semi-major axis evolution, while right show the mutual inclination between the two planets. Top panels place the outer companion at 0.2 au; the system is only weakly perturbed by hot Jupiter destruction. Bottom panels use an external companion at ${\sim}$0.04 au; the system quickly goes unstable after destruction of the inner hot Jupiter. We find that instability requires very close-in companions.}
    \label{fig:stable-and-unstable}
\end{figure*}

In the subsequent analysis, we classify a system as unstable to RLO if either one or more planets is ejected from the system ($e{>}1$), or the hot Jupiter's orbit deviates chaotically from the orbital evolution expected under mass transfer alone (Equation \ref{equation:tracks}). We screen for chaotic deviation from orbital evolution solely under mass transfer by tracking $|a_{\text{sim}}{-}a_{\text{track}}|$ over time; if $\max(|a_{\text{sim}}{-}a_{\text{track}}|)$ exceeds 0.001 au, we classify the system as unstable. Although ad hoc, we find that this prescription reliably identifies systems which are chaotic and thus likely to result in an ejection or merger following the end of the integration.

\Cref{fig:stable-and-unstable} showcases two endmembers of the evolution. The lesson of \Cref{fig:stable-and-unstable} is that there is a regime of companion parameters which become unstable shortly after RLO. While we do not implement collision handling in our numerical calculations, the outcome of such instabilities can be estimated using the Safronov number $\Theta{=}v^{2}_{\rm esc}/2v^{2}_{\rm orb}$, with $v_{\rm esc}$ the planet's escape velocity and $v_{\rm orb}$ the orbital velocity \citep[e.g.][]{saf72,pet25}. We estimate that the stellar potential well near the sub-Jovian desert is sufficiently deep that instabilities should always lead to mergers ($\Theta{<}1$). 
As highlighted in \Cref{fig:stable-and-unstable}, instability is only triggered when companions are sufficiently close-in. We elucidate the origin of this critical separation in Section \ref{sec:mechanism}.

\Cref{fig:stable-and-unstable} also illustrates how the secular oscillations in mutual inclination decrease in amplitude and increase in frequency after RLO, due to the change in the inner planet's semi-major axis and increase in stellar spin rate. This results in the system's mutual inclination being fixed at approximately its pre-RLO value. We thus conclude that systems that are cotransiting immediately before RLO will remain cotransiting after RLO (if they do not go unstable). 

\begin{figure*}
    \centering
    \includegraphics[width=\linewidth]{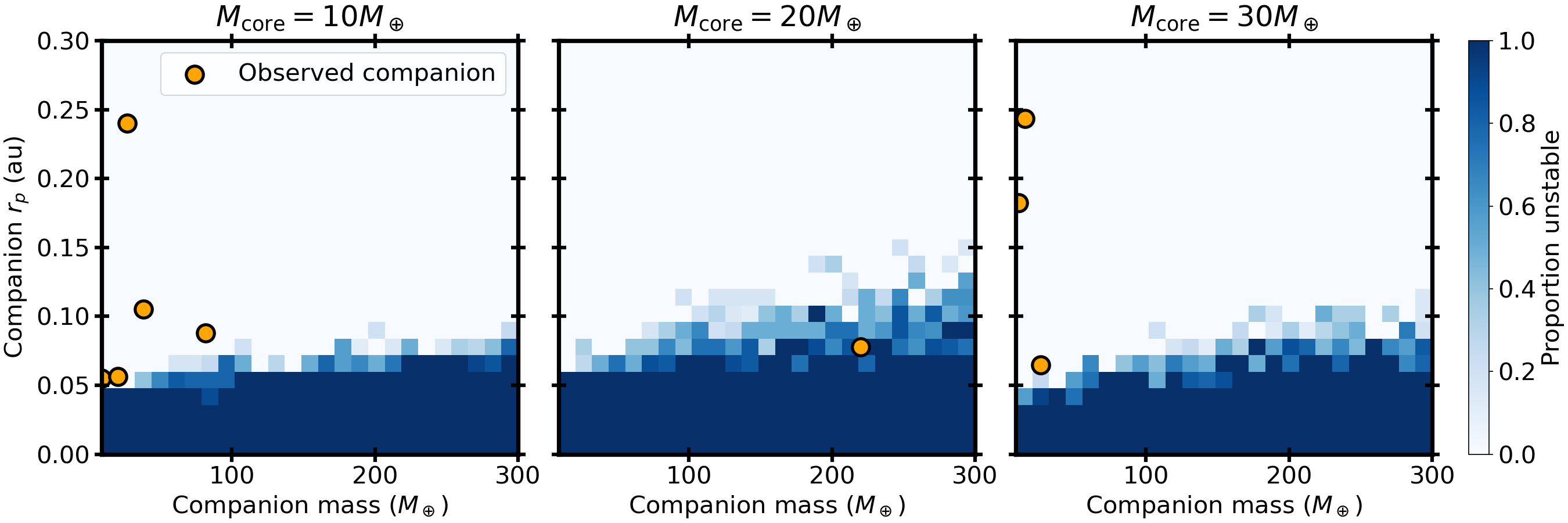}
    \caption{Heatmap recording proportion of synthetic systems that are unstable after the inner hot Jupiter undergoes RLO. Each system consists of an outer companion of varying periapse distance ($r_{p}$, $y$ axis) and mass ($x$ axis), with an inner hot Jupiter that undergoes RLO. Panels use a hot Jupiter of core mass $M_\text{core}{=}10, \ 20, \ 30 \ M_{\oplus}$ from left to right, respectively. Orange markers indicate the parameters of several observed desert dweller companions (see Table \ref{table:summary}), and are placed on the panel with core mass closest to the system's desert dweller mass (we ignore the desert dweller's low-mass atmosphere). The observed companion TOI-1347b (mass $9 M_\oplus$) lies just below the lower bound of our sample $10 M_\oplus$. The instability threshold moves outward with increasing companion mass, hot Jupiter core mass, and final orbital distance of the hot Jupiter remnant (set by its pre-RLO entropy and core mass; see Figure \ref{fig:tracking-test}). Outer companions to desert dwellers must lie at distances ${\gtrsim}$0.05 au (orbital period ${\sim}$4 days) to remain stable; desert dwellers must be alone in the desert if they form via RLO.}
    \label{fig:companion-instability-heatmap}
\end{figure*}

A heatmap of companion masses and periapse distances that yield instabilities is depicted in \Cref{fig:companion-instability-heatmap}. As expected, companions with tighter periapse distances and larger masses are the most likely to undergo instability after RLO. Figure \ref{fig:companion-instability-heatmap} also reveals that the instability threshold in companion distance is set by two effects: hot Jupiter core mass, and the final orbital distance the remnant core reaches after RLO \citep[set by the entropy of the hot Jupiter before RLO; see top left panel of Figure \ref{fig:tracking-test} and][]{halmil26}. Specifically, we observe that the instability threshold for hot Jupiters with 20 $M_{\oplus}$ cores extends to slightly larger companion distances than those with 10 $M_{\oplus}$ cores (inside ${\lesssim}0.1$ vs. ${\lesssim}0.05$ au, respectively). Since hot Jupiters with 10 and 20 $M_{\oplus}$ cores end their RLO at similar orbital distances (for our choice of equal initial planet entropy; see top left panel of Figure \ref{fig:tracking-test}), the differing instability thresholds stem from the difference in core mass; more massive remnant cores incite more instability. On the other hand, using a more massive 30 $M_{\oplus}$ core produces a slightly closer-in instability threshold than a 20 $M_{\oplus}$ core due to its tighter final orbital distance (using the same pre-RLO entropy; see again Figure \ref{fig:tracking-test}). Regardless of initial hot Jupiter entropy/core mass, we observe a global lower limit on companion periapse distance ${\sim}0.05$ au below which systems cannot remain stable. We thus conclude that the orbital period of companions to desert dwellers must exceed ${\gtrsim}$4 days; desert dwellers must be alone in the desert \citep[defined ${\lesssim}$3 days; e.g.][]{casboulil24} if they form via RLO.

We now confirm that the instability limit uncovered in \Cref{fig:companion-instability-heatmap} is truly due to the effect of RLO - i.e. that systems below the threshold would be stable in the absence of RLO. A simple (and convenient to evaluate) diagnostic for circular, two-planet orbital stability requires that they are separated by at least ${\gtrsim}4$ mutual Hill radii \citep[e.g.][]{gla93,wuzhazho19}:

\begin{equation}\label{equation:Khill}
    \frac{a_{2}-a_{1}}{R_{\rm H}}>4
\end{equation}

\noindent with $a_{1(2)}$ planet 1(2)'s semi-major axis, and the mutual Hill radius defined via,

\begin{equation}
    R_{\rm H}=\frac{a_{1}+a_{2}}{2}\bigg(\frac{m_{1}+m_{2}}{3 M_{\star}}\bigg)^{1/3}.
\end{equation}

\noindent Evaluating Equation \ref{equation:Khill} with $m_{1}{=}m_{2}{=}20 \ M_{\oplus}$ (typical desert dweller and companion masses; see Table \ref{table:summary} though we note that this calculation is not sensitive to companion mass), $a_{1}=0.025$ au (a typical desert dweller orbital distance), $a_{2}{=}0.05$ au (at our stability limit), we find a separation ${\sim}20 \ R_{\rm H}$. The two-body stability threshold of Equation \ref{equation:Khill} is only crossed for companion separations $a_{2}{\lesssim}0.03$ au (period ${\sim}$1.9 days). This simple check indicates that two-planet stability alone does not preclude planets in the desert from having other desert-dwelling companions.

We further confirm that neighboring planets in the desert are indeed stable in the absence of RLO using the more accurate chaos indicator derived by \cite{hadlit18}, which can treat massive planets on eccentric orbits. Computing the stability threshold following their Equations 16 and 19, we find that $20 \ M_{\oplus}$ planet pairs at $a_{1}{=0.025}$ and $a_{2}{=}0.05$ au are stable for eccentricities $e_{1(2)}{<}0.7$ for $e_{2(1)}{=}0$ and $e_{1(2)}{<}0.35$ for $e_{2(1)}{=}0.35$ (in the most unstable configuration with longitudes of perihelion $\varpi_{1}{\sim}{\pi}$ and $\varpi_{2}{\sim}0$, respectively). Varying outer companion separation, we find separations $a_{2}{\lesssim}{0.03}$ au produce instability unless both orbits are circular, in line with the threshold estimated via equation \ref{equation:Khill}. We observe a weak dependence of these boundaries on the planet masses. We thus conclude that RLO alone clears out the desert of companions.

\subsection{Mechanism of Instability}\label{sec:mechanism}

The estimates of Section \ref{subsec:singles} indicate that, without RLO,  companions inside our numerically-measured instability threshold would remain stable. Our goal in this section is to elucidate why RLO moves the instability threshold beyond that required for planets to be separated by ${\gtrsim}$a few mutual Hill radii. We determine that mean motion resonance encountered during the hot Jupiter's outward migration pumps eccentricities, to the degree that the system becomes Hill unstable.

\begin{figure}
    \centering
    \includegraphics[width=\linewidth]{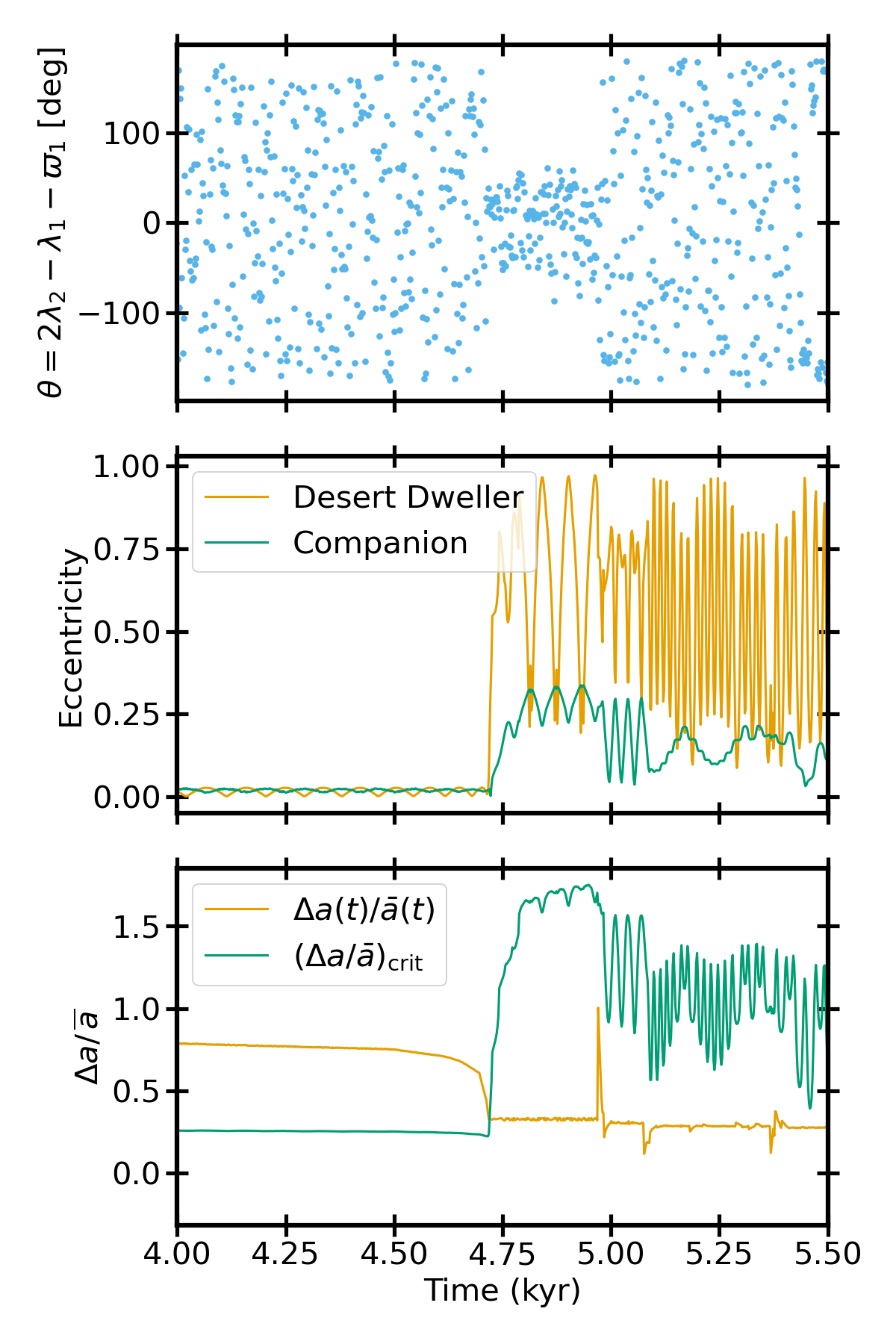}
    \caption{A representative example of orbital instability caused by RLO, with $M_{\text{core}}=10 M_\oplus$. Top to bottom panels: resonant argument associated with the 2:1 commensurability, eccentricity of desert dweller (orange) and companion (green), and planet/planet orbital separation normalized by the mean ($\Delta a/\bar{a}{=}2(a_{2}{-}a_{1}){/}(a_{2}+a_{1})$) for the planet pair (orange) and the threshold for Hill stability \citep[$(\Delta a/\bar{a})_{\rm crit}$, in green; following Equation G.16 of][]{tre23}. The desert dweller's orbital expansion during mass loss establishes a mean motion resonance, pumping eccentricities to the degree that the planets become Hill unstable. We note that, while our \texttt{REBOUND} calculations successfully track the onset of this instability, we do not attempt to follow the planets' evolution afterward (which would require collision handling, a model of tides, etc.)
    \label{fig:mmr-timeseries}}
\end{figure}

An illustration of the instability mechanism is displayed in Figure \ref{fig:mmr-timeseries}. To determine how RLO produces instability, we first analyze the period ratio distribution of our unstable systems shortly before instability (near the end of RLO). We find that these near-instability systems clearly exhibit preference for low-order mean motion resonances, particularly the 2:1. Figure \ref{fig:mmr-timeseries} explicitly confirms in an example system that resonance is achieved, as the critical resonant argument associated with the 2:1 commensurability transitions from circulation to libration. Once resonance is established, eccentricities are pumped. To check whether eccentricities grow enough for the system to become unstable, we compare the planet/planet separation to the Hill stability criterion, corrected for non-zero eccentricity \citep[given by Equation G.16 of][]{tre23}. Boasting eccentricities ${\sim}0.1{-}0.8$ during this resonant phase, we verify that such systems are indeed Hill unstable and (by definition) cannot avoid close encounters. 

We verify that this resonant instability is indeed the primary cause of the instability threshold in our simulations two different ways. First, the period distribution of near-instability systems exhibits clear pileups at low-order ratios, indicating that the majority of our unstable systems achieve resonance before the end of our simulation. Second, the instability boundary from \Cref{fig:companion-instability-heatmap} encompasses the 3:2 to 3:1 range in period ratios, strong evidence that mean motion resonance is responsible for the instability threshold's location. Planets beyond the instability threshold lie far enough away from the desert dweller that they do not encounter low-order mean motion resonance.

\subsection{Effect of Stellar Spin-Up} 

Our calculations include the effect of stellar oblateness due to tidal spin up, but we do not observe that oblateness-induced orbital perturbation plays a significant role in the dynamical evolution. This is somewhat surprising given that our problem resembles the inverse of that explored by \cite{batbod16}, who explored \textit{in situ} formation of hot Jupiters and found that the stellar $J_{2}$ may govern systems containing an inner hot Jupiter and outer companion. In \cite{batbod16}, a hot Jupiter grows quickly in the presence of an outer companion and a rapidly spinning host star. A secular resonance is established as the nodal regression of the hot Jupiter declines due to stellar spin down, sweeping into commensurability with the outer orbit's increasing regression rate due to the growing hot Jupiter. We briefly comment on why our setup -- in which an inner hot Jupiter loses mass and the star spins up -- does not encounter such a resonance. 

We find that, despite the fact that the stellar quadrupole moment grows significantly after RLO (Figure \ref{fig:tracking-test}), the nodal regression rate ${\propto}a^{-7/2}J_{2}$ \citep[][]{murder99,tre23} $\textit{declines}$ due to orbital expansion. As a result of the decline in quadrupolar nodal regression rate, nodal regression for systems with massive (giant planet) external companions is dominated solely by planet/planet secular forcing. Frequency crossings do not occur. 

On the other hand, frequency crossings can occur in systems with lower mass companions (e.g. for companions ${\lesssim}20 \ M_{\oplus}$ at ${\lesssim}0.2$ au). In these cases, the desert dweller nodal regression rate declines into commensurability with the planet/planet regression rate, whose absolute value grows as their orbital separation shrinks \citep[][]{murder99}. We observe however that such crossings are extremely brief, occurring over a short fraction of the hot Jupiter destruction timescale (which itself  is ${\lesssim}10^{4}$ yr). Since such frequency crossings are much faster than the secular oscillation period, we suggest that they occur too rapidly for adiabatic inclination resonance to occur. This rapid frequency crossing is the primary reason why we do not observe the behavior outlined in \cite{batbod16}.

{}

\subsection{Application to Observed Systems}\label{sec:application}

\begin{figure}
    \centering
    \includegraphics[width=\linewidth]{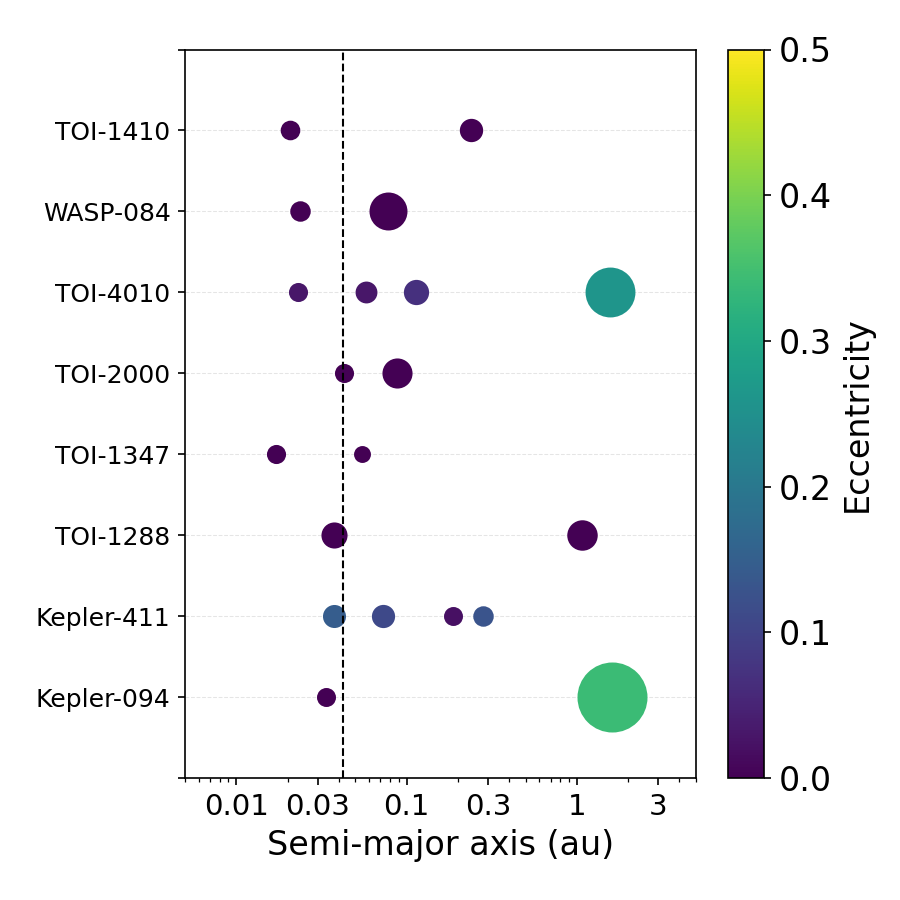}
    \caption{Visual summary of observed desert dweller systems with exterior companions. Each marker area is proportional to the mass of the corresponding planet, with data/references recorded in Table \ref{table:summary}. The vertical dashed line indicates orbital period $P{\sim}3.2$ days, roughly delineating the outer edge of the sub-Jovian desert \citep[e.g.][]{casboulil24}.}
    \label{fig:hn-systems}
\end{figure}

Table \ref{table:summary} and \Cref{fig:hn-systems} summarize systems from the literature containing desert dwellers with exterior companions. We first limit our analysis to observed two-planet systems, corresponding directly to our two-planet $N$-body calculations. We therefore exclude TOI-4010 and Kepler-411 from this initial analysis. We situate the two-planet systems relative to our numerical instability thresholds in \Cref{fig:companion-instability-heatmap}. To assess stability, we plot each system's companion mass and pericenter distance on the panel of \Cref{fig:companion-instability-heatmap} with hot Jupiter core mass closest matching the system's desert dweller mass (ignoring the desert dweller's negligible atmosphere mass). We find that 4/6 observed companions to desert dwellers lie within the stable region of parameter space. The two exceptions are TOI-1347 b and WASP-084 c, with companion parameters on the cusp of the unstable region. The data is therefore consistent with our stability threshold; none of the systems lie in the unambiguously unstable region. One limitation of our estimate however is that, while the observed systems all have stellar masses $0.7{\lesssim}M_{\star}{\lesssim}1 \ M_{\odot}$, our numerical stability threshold is derived using $M_{\star}{=}1 \ M_{\odot}$. Without performing bespoke $N$-body simulations for each system, our conclusions are therefore indicative of stability but are not definitive.

\begin{table*}[]
\centering
\begin{tabular}{l|l|l|l|l|l|l}
\hline
Planet name & desert dweller mass $(M_\oplus)$ & desert dweller $a$ (au) & $N$ & companion mass $(M_\oplus)$ & companion $r_p$ (au) & RLO stable? \\ \hline
Kepler-094b  & $10.8\pm1.4$                & $0.031\pm 0.001$      & 2   & $3126\pm200$                        & $1.04\pm0.01$                 & Yes         \\
Kepler-411b & $25.6\pm2.6$                 & $0.0375\pm0.0008$      & 4   & $26.4\pm 5.9$                        & $0.065 \pm 0.001$                & Yes        \\
TOI-1288b   & $44.1\pm 2.7$                 & $0.0374\pm 0.0008$      & 2   & $85.7\pm 12.1$                          & $1.07\pm 0.03$                 & Yes         \\
TOI-1347b   & $11.1 \pm 1.2$                 & $0.0171\pm 0.0003$      & 2   & $<9.0$                         & $0.0545\pm 0.001$                & Marginal    \\
TOI-2000b   & $11.0\pm 2.4$                 & $0.04271\pm 0.0008$      & 2   & $81.7\pm 4.7$                        & $0.082\pm 0.003$                & Yes         \\
TOI-4010b   & $11.0\pm 1.3$                 & $0.0229\pm 0.0002$      & 4   & $20.31 \pm 2.13$                       & $0.056\pm 0.002$                & Yes\\
WASP-084c   & $15.2\pm 4.5$                 & $0.0236 \pm 0.001$      & 2   & $220\pm 18$                         & $0.0778\pm 0.002$                & Marginal    \\
TOI-1410b    & $12.5\pm1.1$                 & $0.0207\pm 0.0004$      & 2   & $27\pm 3.3$                          & $0.239\pm0.005$                & Yes        
\end{tabular}
\caption{Summary of observed systems containing a desert dweller with exterior companions, and the verdict of our RLO stability analysis. Listed companion parameters are for the inner-most companion in the case of $N{>}2$. Systems with a single companion are classified as RLO stable if they lie within the stable region of \Cref{fig:companion-instability-heatmap}, and marginal if they lie on the boundary between the stable and unstable region. Both systems with $N{>}2$ are inferred to be RLO stable from the simulations described in Section 3.3.} References for each system: Kepler-094: \cite{2024ApJS..270....8W, 2014ApJS..210...20M}; Kepler-411: \cite{2019A&A...624A..15S, 2024ApJS..270....8W}; TOI-1288: \cite{2024ApJS..272...32P}; TOI-1347: \cite{2024AJ....167..153R, 2024ApJS..272...32P}; TOI-2000: \cite{2023MNRAS.524.1113S}; TOI-4010: \cite{2023AJ....166....7K}; WASP-084: \cite{2023MNRAS.525L..43M}; TOI-1410: \cite{2024ApJS..272...32P}.
\label{table:summary}
\end{table*}

For each system, we also verify the long-term stability of the hypothesized initial conditions. The initial conditions we hypothesize consist of the observed system architecture with a hot Jupiter at 0.017 au (the initial condition in our $N$-body runs) replacing the desert dweller. While the hot Jupiters' initial orbital locations are uncertain, we do not expect this choice of initial orbital location (chosen to avoid prohibitively long $N$-body integration run times during the tidal decay phase of RLO) to significantly alter our estimate of pre-RLO stability; to undergo tidal inspiral, the hot Jupiters must begin within a factor ${\sim}2$ of this choice \citep[periods ${\lesssim}$3 days; assuming a stellar tidal quality factor $Q_{\star}{\geq}10^{4}$ as in][]{halmil26}. We verify stability in two ways. First, we evolve each proto-desert dweller system with the \texttt{IAS15} integrator to screen for instabilities \citep{rein2015ias15, rein2012rebound}. We find that each system is stable for at least 1 Myr, corresponding to ${>}10^9$ orbits of the inner hot Jupiter. Furthermore, we use the deep-learning feature classifier \texttt{SPOCK} and find that all 8 systems are inferred to be stable in the long-term \citep{tamayo2020predicting}. Second, we computed the mutual Hill radius following Equation \ref{equation:Khill} for each hypothetical inner hot Jupiter + outer companion system. We confirm that the inner hot Jupiter is separated from outer companions by ${\gtrsim}15 \ R_{\rm H}$, further indicating stability.

We next consider the stability of the two systems with more two planets: TOI-4010 and Kepler-411. We carry out two additional sets of integrations to model systems with higher multiplicity. Specifically, we choose the observed external companion parameters of the TOI-4010 and Kepler-411 systems. Our integrations reveal that both the TOI-1410 and Kepler-411 systems remain stable after RLO. We also verified the stability of the initial conditions, consisting of a hot Jupiter at 0.017 au replacing each system's desert dweller, by directly integrating system evolution over ${\sim}10^7$ orbits of the inner hot Jupiter. We find that the systems are indeed stable, and thus compatible with an initial architecture containing an inner hot Jupiter.

We therefore conclude that all but two observed multiplanet desert dweller systems are dynamically stable through RLO. The two exceptions sit on the boundary of (in)stability. 

\section{Other Tests $\&$ Future Work}\label{sec:future}

While this Letter focuses on the existence of companions to discern between HEM and tidal inspiral as desert dweller formation theories, additional axes of differentiation should be explored. We briefly highlight two additional channels of differentiation future work should address.

One clean difference in prediction between the HEM and tidal inspiral pictures concerns the sub-Jovian desert's stellar type dependence. In the \cite{halmil26} theory, the sub-Jovian desert is expected to remain empty around hot stars above the Kraft break \cite[][]{kra67}. This expectation stems from the fact that such hot stars are thought to yield weaker tidal dissipation than Sun-like stars, as hot stars do not possess thick convective envelopes (unlike Sun-like stars) and experience weaker magnetic braking than cool stars \citep[e.g.][]{winfabalb10,albwinjoh12}. This weaker tidal dissipation likely precludes tidal inspiral and subsequent RLO. On the other hand, tidal disruption of hot Jupiters at very high eccentricity should occur indiscriminately above and below the Kraft break. To our knowledge, the sub-Jovian desert currently does not contain any planets around stars above the Kraft break; future statistical work correcting for biases would be helpful to ensure that this is a real feature of the planet population. If so, stellar type dependence in tandem with the existence of companions offer a powerful joint constraint on theories of the sub-Jovian desert.

Another clear difference between HEM and tidal inspiral concerns the stellar obliquities of planets in the desert. Similar to hot Jupiters, desert dweller stellar obliquities should initially be excited if they are emplaced via HEM. Conversely, tidally-driven RLO tilts stellar obliquities into alignment (within ${\sim}$a few tens of degrees; Hallatt, Owen $\&$ Millholland, subm.). Follow-up obliquity measurements in the sub-Jovian desert would therefore also be helpful.

\section{Conclusion}\label{sec:conclusion}

Competing theories for 
the origin of planets residing in the sub-Jovian desert make differing predictions for the existence of neighbors.  If desert dwellers are the products of gas giants destroyed via HEM, they should be strictly alone \citep[][]{musdavjoh15, bec25}. If instead desert dwellers form via hot Jupiter tidal inspiral, as suggested by \cite{halmil26}, a fraction of desert dwellers may inherit the nearby companions of hot Jupiters \citep[e.g.][]{shavanhua26}. The goal of this Letter is to determine whether external companions of hot Jupiters remain dynamically stable as the hot Jupiter is destroyed via RLO. 

We have shown that RLO clears out the desert of companions with periods inside ${\sim}4$ days; desert dwellers should be alone in the desert if they form via RLO. Mapping the companion stability threshold across companion mass, orbital distance, and desert dweller mass, we find that 6/8 observed systems containing a desert dweller with companions are stable through RLO. The remaining two systems sit on the cusp of instability. 
As the desert dweller sample continues to grow \citep[e.g.][]{carcaspal26}, our analysis indicates that those with companion planets outside the desert will remain consistent with the \cite{halmil26} mechanism. Future observations estimating the occurrence rate of desert dwellers with companions will therefore help discern between desert dweller formation theories, and by extension, elucidate the fate of hot Jupiters.

\section{Acknowledgments}
We thank the anonymous referee for helpful comments that improved our paper considerably. We also thank Daniel Yahalomi for constructive feedback. This material is based upon work supported by the National Science Foundation under Grant No. 2306391. The authors acknowledge the MIT Office of Research Computing and Data and the MIT Engaging Cluster for providing computational resources that contributed to the results reported in this paper.

\bibliography{biblio}{}
\bibliographystyle{aasjournal}

\end{document}